\documentclass[a4paper,usenatbib,useAMS]{mnras}
\usepackage{graphicx}	
\usepackage[flushleft]{threeparttable}
\usepackage{amsmath}	
\usepackage{amssymb}	
\usepackage{multicol}        
\usepackage{bm}		
\usepackage{pdflscape}	
\usepackage{color}
\usepackage{hyperref}
\usepackage{natbib}
\usepackage[T1]{fontenc}
\usepackage{ae,aecompl}
\usepackage{newtxtext,newtxmath}
\usepackage{xcolor}
\usepackage[utf8]{inputenc} 
\newcommand{\kms}{\,km\,s$^{-1}$} 
 
\definecolor{darkspringgreen}{rgb}{0.09, 0.45, 0.27}
\def\Hnot{\mbox{$H_{0}$ }}

\title[Spectroscopy and Models of 2 lensed quads in DES]{Spectroscopic confirmation and modelling of two lensed quadruple quasars in the Dark Energy Survey public footprint}

\author[C.~Spiniello et al.] {C.~Spiniello$^{1,2}$, A.~V.~Sergeyev$^{3,4}$, L.~Marchetti$^{5,6,7}$, C.~Tortora$^{8}$, N.~R.~Napolitano$^{1,9}$, \and V.~Shalyapin$^{10,11}$, A.~Agnello$^{12}$, F.~I.~Getman$^{1}$, M.~Vaccari$^{6,7}$, S.~Serjeant$^{13}$, \and  L.~V.~E.~Koopmans$^{14}$, A.~J.~Baker$^{15}$,  T.~H.~Jarrett$^{5}$, G.~Covone$^{16,17}$, G.~Vernardos$^{14}$ \\
$^{1}$INAF - Osservatorio Astronomico di Capodimonte, Salita Moiariello, 16, I-80131 Napoli, Italy \\
$^{2}$European Southern Observatory, Karl-Schwarschild-Str. 2, 85748 Garching, Germany \\
$^{3}$Astronomical Institute of Kharkov National University, 61022, 35 Sumskaya St, Kharkov, Ukraine \\
$^{4}$Institute of Radio Astronomy of the National Academy of Sciences of Ukraine \\
$^{5}$ Department of Physics and Astronomy, University of Cape Town, Private Bag X3, Rondebosch 7701, Cape Town, South Africa \\
$^{6}$Department of Physics and Astronomy, University of the Western Cape, Private Bag X17, Bellville 7535, Cape Town, South Africa \\
$^{7}$INAF - Istituto di Radioastronomia, via Gobetti 101, 40129 Bologna, Italy\\
$^{8}$INAF - Osservatorio Astrofisico di Arcetri, Largo Enrico Fermi 5, 50125, Firenze, Italy \\
$^{9}$School of Physics and Astronomy,  Sun Yat-sen University Zhuhai Campus, 2 Daxue Road,  Tangjia,  Zhuhai,  Guangdong 519082,  P.R. China \\
$^{10}$O. Ya. Usikov Institute for Radiophysics and Electronics, National Academy of Sciences of Ukraine, 12 Ac. Proskura St., UA-61085 Kharkov, Ukraine \\
$^{11}$Departamento de F\'isica Moderna, Universidad de Cantabria, Avda. de Los Castros s/n, E-39005 Santander, Spain\\
$^{12}$DARK, Niels Bohr Institute, Copenhagen University, Lyngbyvej 2, 2100 Copenhagen, Denmark\\
$^{13}$ School of Physical Sciences, The Open University, Milton Keynes MK7 6AA, UK \\
$^{14}$ Kapteyn Astronomical Institute, University of Groningen, PO Box 800, 9700 AV Groningen, the Netherlands\\
$^{15}$ Rutgers, The State University of New Jersey, Department of Physics
and Astronomy, 136 Frelinghuysen Road, Piscataway, NJ 08854-8019, USA\\
$^{16}$ Dipartimento di Fisica 'E. Pancini', Universit\'a di Napoli Federico II, Compl. Univ. Monte S. Angelo, I-80126 Napoli, Italy\\
$^{17}$ INFN, Sezione di Napoli, Compl. Univ. Monte S. Angelo, I-80126 Napoli, Italy}
\date{Last updated 2018 December 22}

\pubyear{2019}
\begin{document}
\label{firstpage}
\maketitle

\begin{abstract}
Quadruply lensed quasars are extremely rare objects, but incredibly powerful cosmological tools. 
Only few dozen are known in the whole sky.  
Here we present the spectroscopic confirmation of two new quadruplets WG0214-2105 and WG2100-4452 discovered by \cite{Agnello18} within the Dark Energy Survey (DES) public footprints. 
We have conducted spectroscopic follow-up of these systems with the Southern African Large Telescope as part of a program that aims at confirming the largest possible number of optically selected strong gravitational lensing systems in the Equatorial and Southern Hemisphere. 
For both systems, we present the spectra for the sources and deflectors that allowed us to estimate the source redshifts and unambiguously confirm their lensing nature. For the brighter deflector (WG2100-4452), we measure the stellar velocity dispersion from the spectrum. We also obtain photometry for both lenses, directly from DES multi-band images, isolating the lens galaxies from the quasar images. One of the quadruplets, WG0214-2105, was also observed by Pan-STARRS, allowing us to estimate the apparent brightness of each quasar image at two different epochs, and thus to find evidence for flux variability. This result could suggest a microlensing event for the faintest components, although intrinsic variability cannot be excluded with only two epochs. Finally, we present simple lens models for both quadruplets, obtaining Einstein radii, SIE velocity dispersions, ellipticities, and position angles of the lens systems, as well as time delay predictions assuming a concordance cosmological model. 
\end{abstract}


\begingroup
\let\clearpage\relax
\endgroup
\newpage

\section{Introduction} 
\label{sec:intro}
According to the estimate of \cite{Oguri10}, quadruplets represent $\sim14\%$ of a statistically-complete sample of lensed quasars. The quadruplet configuration, in fact, requires a closer alignment between the source and the deflector than the one needed for the more common doublet configuration\footnote{We note however that the magnification bias \citep{Turner80}, due to the fact that multiple imaging magnifies the source, is larger in the case of a quadruplet configuration, compensating for the smaller cross-section.}. 
However, when the number of source images is doubled, the information that can be gathered in terms of stellar mass fraction and deflector potential is larger (e.g.,  \citealt{Schechter04, Treu16}). 
Quadruplets have also been shown to be valuable cosmological tools \citep{Refsdal64, Blandford92, Witt00, Suyu13, Treu16, Bonvin17} and useful for microlensing studies to study the inner parts of quasar accretion discs \citep{Schechter02, Kochanek04, Eigenbrod08, Blackburne11, Mediavilla17, Vernardos18}.  
For instance, studies based on blind analysis have shown that a single quadruplet can be used to measure the so-called time delay distance ($D_{\Delta t}$), a multiplicative combination of the three angular diameter distances between the observer, deflector and source \footnote{$D_{\Delta t}=(1+z_{l})[D_{l}D_{s}/D_{ls}]$, with $z_{l}$ the redshift of the deflector, $D_{l}$ the angula diameter distance of the lens from the observer, $D_{s}$ the distance of the source from the observer and $D_{ls}$ the relative distance between the lens and the source.}, with an uncertainty of 5-7\% \citep{Suyu12}. 
Since $D_{\Delta t}$ is inversely proportional to the Hubble constant, \Hnot  \citep{Linder11,Suyu12}, and more weakly dependent on other cosmological parameters, obtaining a precise estimation of it through time-delay lenses allows one to break some of the main degeneracies in the interpretation of cosmic microwave background data . 
Moreover, time delay distances are independent of the local distance ladder and give comparable precision on \Hnot, providing a crucial test of potential systematic uncertainties\footnote{See, e.g., the work done by the \Hnot  Lenses in COSMOGRAIL's Wellspring (H0LiCOW, https://shsuyu.github.io/H0LiCOW/site/) Collaboration, 
of which some of the authors of this publication are members.}.

In this paper, we report the spectroscopic confirmation of two recently discovered quadruplets: WG0214-2105 and WG2100-4452, both identified in the Dark Energy Survey (DES, \citealt{Abott18}) DR1 public footprints, using the source-based methods developed and presented in \citet[hereafter A18]{Agnello18} to find lens candidates in wide-sky photometric multi-band surveys.  The spectroscopic confirmation has been carried out using the Southern African Large Telescope (SALT; \citealt{Buckley06}) as part of the observing program \textit{Gotta catch'em all} (2018-2-SCI-020, PI: Marchetti). 

\medskip

The paper is organized as follow. 
In Section~\ref{sec:SALT}, we give details on the SALT observing program and the selected candidates. A brief description of the two new quadruplets, including their coordinates and infrared magnitudes used at the pre-selection stage, is provided in Section~\ref{sec:quads}. 
In Section~\ref{sec:photometry}, we report the results of a photometric analysis, which then leads to a preliminary variability analysis for one of the systems.  
In Section~\ref{sec:spectroscopy}, we present the obtained SALT spectra for both systems, and highlight the redshift determinations for the sources and the deflectors. In Section~\ref{sec:models}, we provide simple lensing models for both quadruplets. 
We summarize our findings and conclude in Section~\ref{sec:conclusions}.  

Throughout the paper, we adopt a flat $\Lambda$CDM cosmology with $\Omega_M=0.3$, $\Omega_{\Lambda}=0.7$ and $\Hnot=70$ kms$^{-1}$Mpc$^{-1}$. 

\begin{table*}
\caption{Coordinates, infrared magnitudes from the WISE catalog \citep{Wright10}, and spectroscopic observations settings for the two quadruplets. For WG0214-2105 we obtained two different observation blocks, with two different position angle orientations (-35 and +50, degrees respectively), as indicated in the table.}
\label{tab:systems}
\begin{center}
\begin{tabular}{|c|c|c|c|c|c|c|c|c|c|c|}
\hline 
\hline 
ID & RA & Dec & W1  & W2  & W3  & W4 &  Total Exp. & Slit  & PA  \\ 
 & (J2000) & (J2000) &  (mag) & (mag) & (mag) & (mag) & Time(sec) & Width ($\arcsec$) & (deg.)  \\
\hline 
WG0214-2105 & 02:14:16.4 & -21:05:35.3 & $15.11\pm0.03$ & $14.67\pm0.06$ & $11.8\pm0.2$ & $8.6\pm0.3$ & 1800 & 2.0 & $-35/+50$  \\
WG2100-4452 & 21:00:14.9 & -44:52:06.4 & $14.14\pm0.03$ & $13.42\pm0.03$ & $10.7\pm0.1$ & $8.2\pm0.3$ & 1500 & 3.0 & +40 \\
\hline 
\hline 

\end{tabular} 
\end{center}
\end{table*}

\section{The SALT Observing Program} 
\label{sec:SALT}
Our observing pprogram: \textit{"Gotta catch’Em All}, the spectroscopic follow-up of strong gravitational lenses from KiDS and KABS surveys" (PI: L. Marchetti, ID: 2018-2-SCI-020) has the goal of spectroscopically confirming optically selected strong gravitational lens (SL) candidates, both quasi-stellar object (QSO)-galaxy (\citealt{Spiniello18}, hereafter S18 and \citealt{ Agnello18}, hereafter A18), and galaxy-galaxy (gravitational arcs, \citealt{Petrillo18, Petrillo19}, hereafter P18 and P19 respectively) systems. 
The candidates have been found by exploiting the improved image quality and the more extended and homogeneous sky coverage achieved with new, deep optical surveys. 
In particular, we focus on the Kilo Degree Survey (KiDS, \citealt{deJong15, deJong17}), because of its depth, exquisite image quality and quite stringent seeing constraints (limiting magnitude of 25 at $5\sigma$ in $2\arcsec$ aperture and mean FWHM of $0.7\arcsec$ in r-band), and on the KiDS VST-ATLAS Bridging Survey (KABS, Napolitano et al., 2019, in prep)\footnote{KABS is a new Guarantee Time Observation program at the VLT Survey Telescope (VST, \citealt{Capaccioli11}), which is equipped with the wide field camera OmegaCAM \citet{Kuijken11} and operating at the ESO observatory in Cerro Paranal (Chile).}, which images a previously almost uncovered region of the southern sky. 
Finally, we also include a few very promising candidates from the DES public footprint, which were recently found by our team \citep{Agnello18}.

After almost two years, we have collected a list of ~300 high-grade candidates  which need spectroscopic confirmation and redshift measurements to translate our lens model results (e.g., Einstein radii) into physical mass measurements.  We have already presented the search methods as well as the KiDS candidates in recently published papers (\citealt{Petrillo17, Petrillo18, Spiniello18, Petrillo19}), and we will present the KABS candidates in a paper in preparation (Spiniello et al., 2019). 

Full results from the first semester of SALT observations, still on-going, will be presented in a dedicated, future, paper (Marchetti, Tortora, Spiniello et al., in prep.). 
In this paper, we focus on the DES candidates: 
two quadruply lensed QSOs for which the spectroscopic confirmation was the last required step to unambiguosly confirm their lensing nature. These pilot observations have been carried out to test our strategy and integration times before proceeding with a larger candidate list (arcs and QSOs) from the KIDS and KABS surveys.

\begin{table*}
\caption{Astrometric and photometric properties of all components of the two systems. Relative positions are computed from the $r$-band DES images and always use  as reference the brightest QSO image (A). Magnitudes in the $griz$ bands have been computed using \texttt{galfit}.}
\label{tab:table_astrometry}
\begin{center}
\begin{tabular}{|c|c|c|c|c|c|c|c|}
\hline
\hline
ID &  comp. & $\delta$x & $\delta$y &  g  & r  &  i  & z  \\ 
&  & ($r$-band, $\arcsec$) & ($r$-band, $\arcsec$) &  (mag) & (mag) & (mag) & (mag)\\
\hline 
WG0214-2105 & A & $+0.000\pm0.005$ &$0.000\pm0.003$ &  $20.47\pm0.04$ & $20.30\pm0.02$ & $20.33\pm0.02$ & $20.14\pm0.02$ \\ 
WG0214-2105  & B & $+0.460\pm0.003$ & $-0.915\pm0.003$ & $20.39\pm0.03$ & $20.26\pm0.02$ & $20.26\pm0.02$ & $20.07\pm0.03$ \\
WG0214-2105  & C & $-0.852\pm0.003$ & $-1.678\pm0.003$ & $20.50\pm0.04$ & $20.44\pm0.04$ & $20.34\pm0.02$ & $20.18\pm0.02$ \\
WG0214-2105  & D & $-0.876\pm0.008$ & $-0.234\pm0.005$ & $21.11\pm0.06$ & $20.83\pm0.03$ & $20.71\pm0.03$ & $20.61\pm0.03$ \\
WG0214-2105  & G & $-0.34\pm0.05$ & $-0.96\pm0.05$ & $20.83\pm0.15$ & $20.00\pm0.07$ & $19.46\pm0.04$ & $19.37\pm0.06$ \\
\hline 
WG2100-4452 & A & $+0.000\pm0.011$  & $0.000\pm0.008$  &  $19.91\pm0.05$ & $19.85\pm0.05$ & $19.58\pm0.10$ & $19.64\pm0.06$\\
WG2100-4452 & B & $-0.437\pm0.005$ & $-0.332\pm0.005$ & $19.23\pm0.02$ & $19.12\pm0.02$ & $18.99\pm0.05$ & $18.84\pm0.02$\\
WG2100-4452 & C & $-2.48\pm0.01$ & $0.36\pm0.01$  & $21.10\pm0.01$ & $21.07\pm0.02$ & $20.82\pm0.04$ & $20.70\pm0.04$\\
WG2100-4452 & D & $-0.86\pm0.02$   & $1.90\pm 0.02$   & $21.70\pm0.03$ & $21.52\pm0.05$ & $21.31\pm0.08$ & $21.20\pm0.06$\\
WG2100-4452 & G & $-1.039\pm0.003$ & $0.823\pm0.005$  &  $18.81\pm0.01$ & $17.68\pm0.01$ & $17.30\pm0.01$ & $16.83\pm0.04$\\
\hline
\hline
\end{tabular} 
\end{center}
\end{table*}

\begin{figure*}
\begin{center}
\includegraphics[scale=0.7,angle=0]{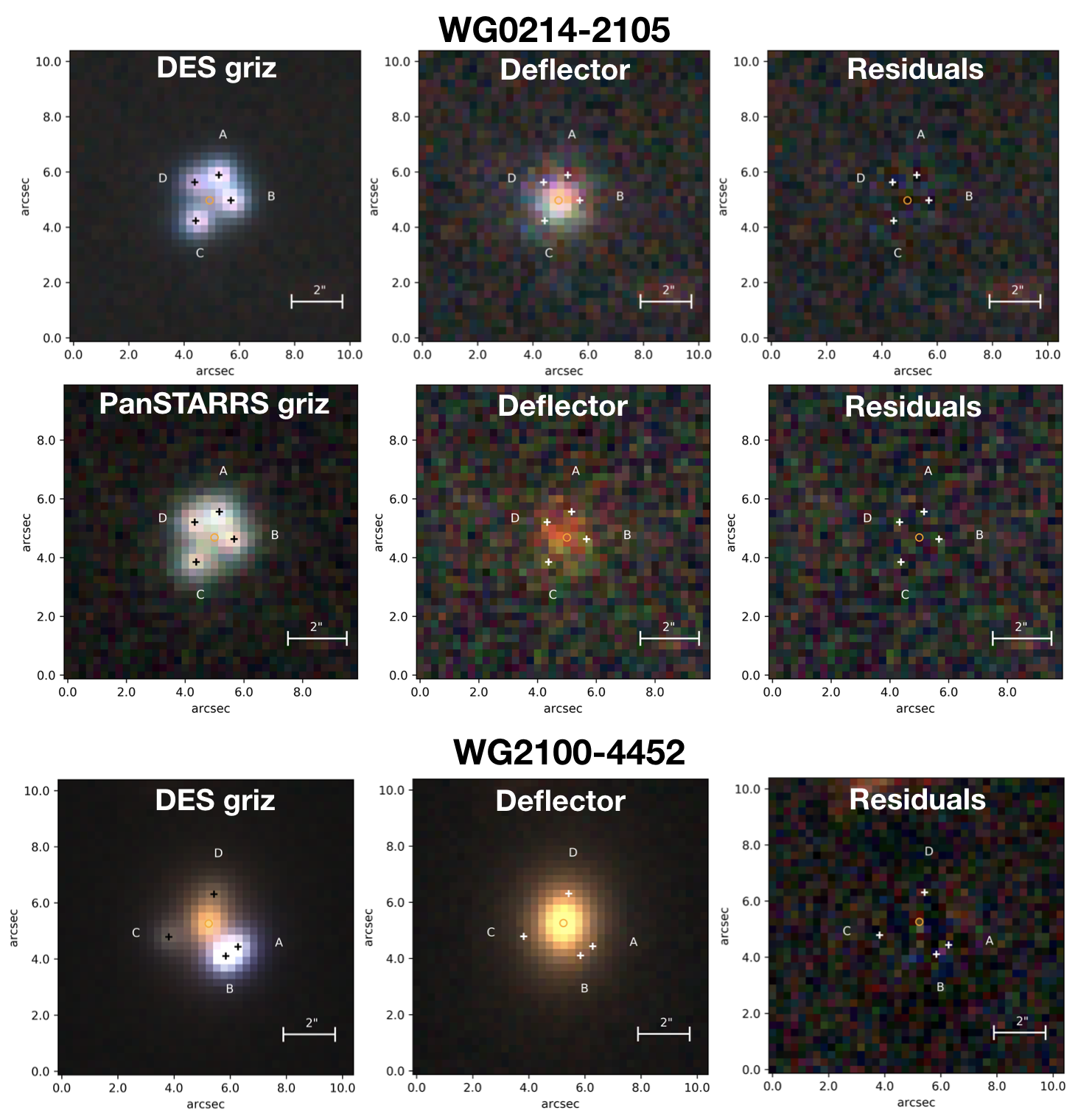}
\caption{Cutouts generated combining $g$-, $r$-, $i$- and $z$-band DES images (top panels) and Pan-STARRS images  (middle panels) for WG0214-2105, and for WG2100-4452 only from DES (bottom panels). For each system, we show the results after subtracting the multiple QSO images (middle column) and the residuals obtained after also subtracting the deflector (right column).}
\label{fig:psf}
\end{center}
\end{figure*}

\section{Two new quadruplets in the Southern Hemisphere: WG0214-2105 and WG2100-4452}
\label{sec:quads}
\citet[hereafter ARN18]{Agnello18RN} reported the discovery of  WG0214-2105 (RAJ2000: 02:14:16.37, DECJ2000: -21:05:35.3), found by cross-matching the publicly available data of three wide-area sky surveys in the southern hemisphere (DES DR1, \citealt{Abott18}; VST-ATLAS, \citealt{Shanks15}; and Pan-STARRS, \citealt{Chambers16}).  The multiple images, in the the typical fold/cross configurations, are characterized by "white" optical colors, "blue" mid-IR colors and high UV deficit (ARN18).  

The discovery of WG2100-4452 (RAJ2000: 21:00:14.9, DECJ2000: -44:52:06.4), has been reported in A18, where we exploit several complementary methods to search for lensed quasars in wide-area photometric surveys. 
In particular, this previously unknown quadruplet in the DES footprint has been found by pre-selecting QSO-like objects based on infrared colors (from the Wide-Infrared Survey Explorer, WISE, \citealt{Wright10}) and then making use of the high spatial resolution of the Gaia Mission (\citealt{GaiaDR16, Lindegren16}) to identify single WISE objects with multiple Gaia matches. We refer the reader to A18 for a more comprehensive description of this and the other methods, and for a complete list of other similarly identified lens candidates. 

Both WG0214-2105 and WG2100-4452 have also been shown to be easily compatible with simple lens models \citep{Wynne18}.

In this paper we present their spectroscopic confirmation, with estimation of the source and deflector redshifts, as well as $griz$-photometry obtained directly from the DES images and lensing modeling. We note that a very recent paper, contemporary to this work, has been published by \cite{Lee19}, reporting a spectroscopic confirmation of WG0214-2105 with Gemini Multi-Object Spectrographs (GMOS) spectra. Here we provide a more detailed analysis of the spectral, photometric and lensing properties of the latter system, comparing our inferences for the source and deflector redshifts with the related findings in \cite{Lee19}.

Table~\ref{tab:systems} reports, for the two lens systems, coordinates and WISE infrared magnitudes, used for pre-selection and given in their native Vega system. 
We also give details about the observational blocks (slit position, orientation and integration times) that we will present in detail in Section~\ref{sec:spectroscopy}. 

\section{Accurate astrometry and PSF Photometry analysis}
\label{sec:photometry}
Following the recipe detailed in S18, we perform Direct Image Analysis (DIA) on the multi-band DES images of the two systems. 
Briefly, we simultaneously fit a point-PSF model to all the QSO multiple images in each of the single-band images, for $griz$ bands, and then generated subtracted images that we visually inspected to identify the  position of the deflector. 
Accurate astrometry is, in fact, of great help when extracting the 1D spectra from the long-slit data and it is also very useful when performing lensing and mass modelling. 

We then derive for each band photometry on all the components (A = brightest QSO image, D = faintest QSO image, G = lens galaxy) simultaneously using the code  \texttt{galfit} \citep{Peng02}. We calculated the Point Spread Function (PSF) from a nearby star, to minimize the effect of distortion on the field and then modelled  each quasar image simultaneously with a point-like PSF profile and the galaxy with a Se\'rsic profile convolved with the PSF. 
Specifically, we assume a de Vaucouleurs profile using a fixed Se\'rsic index\footnote{We note that more detailed modelling, letting for instance the Se\'rsic index varying as free parameter, is not ideal since the images are too noisy.} n=4, to model the light distribution of the lens galaxy.  

Figure~\ref{fig:psf} shows the results of the PSF fitting. For WG0214-2105, we also  perform DIA on the Pan-STARRS images. In each line of the Figure, the left-most panel shows the $griz$-combined color cutouts of  $10\arcsec\times10\arcsec$ size, and the middle panel shows the same image after the 4 QSO components have been subtracted, and from which it is possible to identify the presence of the deflector. Finally, the right-most panel shows the residuals after fitting and subtracting also the light of the lens galaxy. 
Table~\ref{tab:table_astrometry} reports the relative positions of all the QSO components obtained from the $r$-band DES images, as well as the inferred magnitudes for $g$-, $r$-, $i$-, and $z$-band with their uncertainties, as calculated by \texttt{galfit}. 

\subsection{Variability Evidence for WG0214}
\label{sec:microlensing}
Quasar microlensing occurs when the light of the source, already deflected by the lens gravitational field, is also affected by the gravitational field of relatively low-mass ($10^{-6}<m/M_{\odot}<10^{3}$) bodies moving along the line-of-sight (e.g., single stars, brown dwarfs, planets, globular clusters, etc.). 
More specifically, microlensing is observed when the angular size of such bodies is smaller than the Einstein radius ($\theta$) of the lensing system and therefore comparable with quasar angular size.  Thus, microlensing is a very useful tool to constrain quasar structure as well to estimate the masses of these compact bodies \citep{Kochanek04}.

During a microlensing event, the mass of the micro-deflector is not large enough to cause a measurable displacement of the light path, but, due to the relative motion between the source quasars, the lenses and the observer, a change in the magnification with time can be detected \citep{Schmidt10}. Thus, the apparent brightness of the source can change by more than a magnitude on time scales of weeks to months or years\footnote{The first detection of a microlensing event from photometric variations was reported in \citet{Irwin89}, for the Q2237+0305 lens system.}.
Unlike the intrinsic quasar variability, however, microlensing variability is uncorrelated between images.

Microlensing has a number of very important astrophysical applications. In particular, it can help in determining the existence and effects of compact objects along the line-of-sight, in studying in great detail the size, structure and light distribution of the source quasars \citep{Dai10,Morgan10,Odowd15} and in constraining the masses and mass distributions of lensing objects and their environments (e.g., \citealt{Cowley17}). 

\begin{figure}
\begin{center}
\includegraphics[scale=0.44,angle=0]{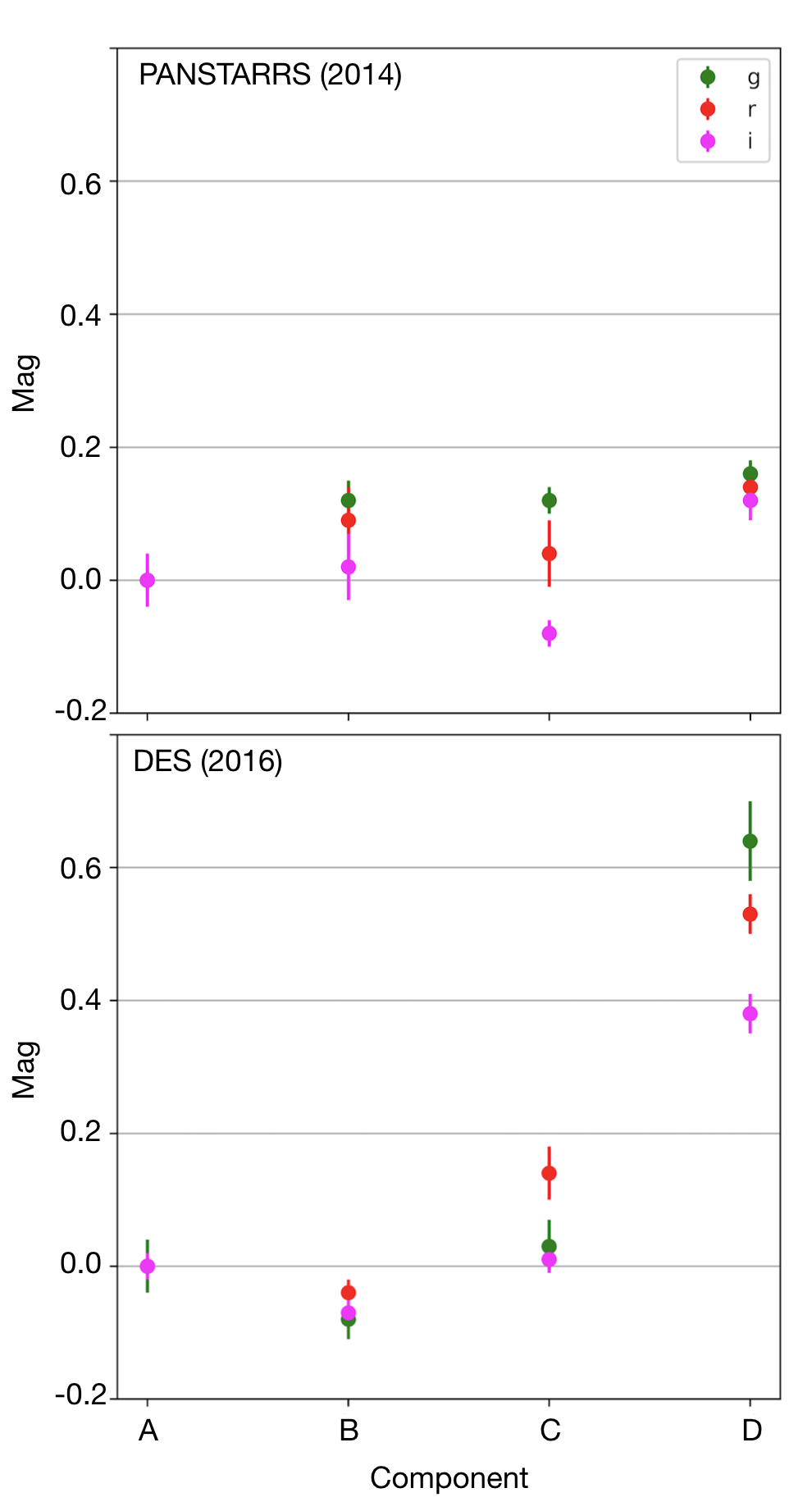}
\caption{Estimated magnitudes at different filter bands ($g$, $r$, $i$) from Pan-STARRS images taken in 2014 (upper panel) and from DES images taken in 2016 (lower panel) for the four QSO images of WG0214-2105. The brightnesses of the B and the D components changes with time, without a clear correlation with wavelength. This result could indicate the presence of microlensing.}
\label{fig:microlensing}
\end{center}
\end{figure}

For WG0214-2105, we can compare two magnitude measurements obtained from multi-band images taken at different epochs: DES, taken in 2016, and Pan-STARRS,  taken in 2014. 
Although with only two measurements one cannot build a detailed light-curve, these data allowed us to detect time variation of the flux of each quasar component, which appears to be uncorrelated with wavelength and thus suggestive of a microlensing origin.  
Figure~\ref{fig:microlensing} plots the magnitudes calculated for different filter bands ($g$, $r$, $i$) from Pan-STARRS (upper panel) and DES (lower panel) for the four QSO images of WG0214-2105. We excluded the $z$-band image because it was very noisy for the Pan-STARRS case.
The brightness of the B, C and D components changes with time, without any clear correlation between wavelength and magnitude. This result seems to indicate a microlensing effect, although we stress that chromatic variability cannot be entirely excluded based on just two epochs of broad-band imaging. 
Another argument in favor of microlensing comes from \citet{Palanque11} and in particular from their Fig.5, which shows the variability structure function for a typical quasar (the change in amplitude as a function of time between observations at different epochs). The function predicts that a change of $\sim0.3-0.5$ mag, similar to what we observe for the D component, is expected only if the two epochs are separated by several years\footnote{We note that optically violent variable (OVV) quasars can vary at this level on time-scales of days; however these systems are extremely rare.}.
We caution the reader, however, that delay among different lightcurves might cause a large flux variation, as shown in \citet{Courbin18} for the lens D0408. Flux ratio variations among lightcurves also have implications for claims of substructure detection, as was shown explicitly by \citet{Agnello17} on the same lens.

\section{SALT Spectroscopic Follow-up}
\label{sec:spectroscopy}
In November 2018, we obtained long-slit spectroscopy for the two quadruplets with the Robert Stobie Spectrograph (RSS) instrument and the PG0900 grating (grating-angle = 15.875), covering from 4490 to 7540 \AA with resolution $R = 1000$ measured from the OI5577 sky line, and using $2\times4$ binning, corresponding to $\sim0.5\arcsec$/pix in the spatial direction and to 0.97\AA pix$^{-1}$ in the dispersion direction. 

As already highlighted, these two DES systems, being bright and visible at the  beginning of the SALT observation semester, have been used as "test-cases" in modest observing condition (seeing$>1.5\arcsec$) and non-optimal observing time (twilight), mainly to test our choices of grating, spatial binning and integration time on target.

The data were reduced with two independent methods, namely the SALT science pipeline (PySALT, \citealt{Crawford10})\footnote{http://pysalt.salt.ac.za/}
and standard \texttt{IRAF} routines. 
One-dimentional (1D) spectra were extracted and analysed with custom \texttt{IDL} and \texttt{Mathematica} codes. 
In the following section we discuss the extraction of the spectra, deblending of the components and estimation of the redshift for each target separately. 

\subsection{WG0214-2105}
We observed WG0214-2105 in two separate observing blocks with a slit of $1.5\arcsec \times 8\arcsec$ and two different position angles (PA): -35 degrees (PA-35 configuration) and +50 degrees (PA+50 configuration). Both observations were centered on the deflector, and both integrated for 1800 seconds on target. Seeing was not ideal ($\approx 1.8\arcsec$) but the night was clear.  
These configurations were motivated by the need to maximize the exposure time on the faint lens galaxy, and simultaneously obtain at least two completely independent spectra of the QSO components (A+B+C from PA-35 and A+C from PA+50).

\begin{figure}
\includegraphics[scale=0.35,angle=0]{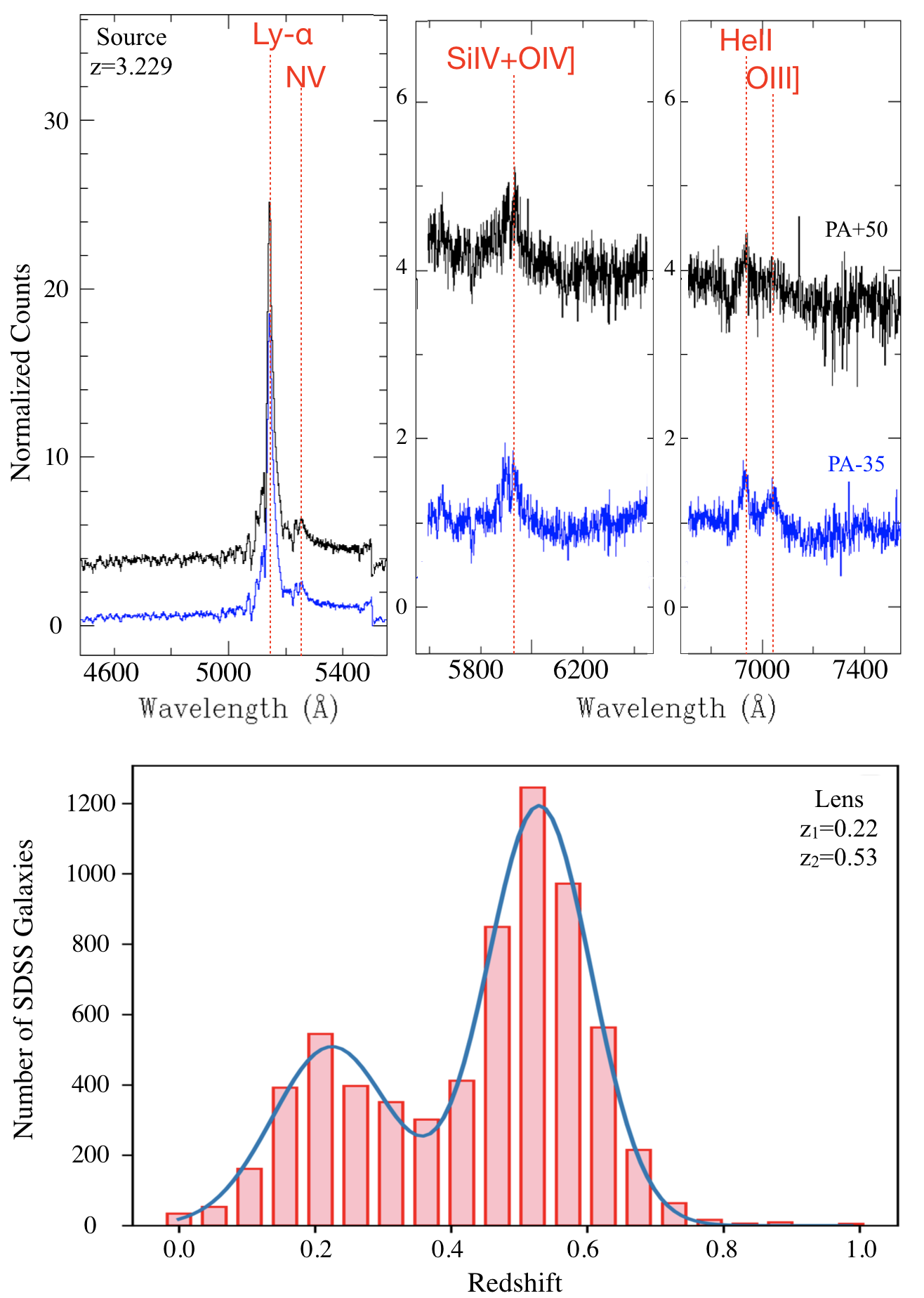}
\caption{SALT spectra of the lensed quasar (upper panels) and redshift distribution of the deflector (lower panel) for WG0214-2105. For the source, we show the two independent spectra extracted from the two slits with different position angles, corresponding to the A+C components (PA+50, black) and the A+B+D components (PA-35, blue). We zoom-in on the most prominent emission lines (Ly-$\alpha$
in the left panel, SiIV+OIV] in the middle panel and HeII and OIII] in the right panel) that allowed us to securely infer its redshift. 
Unfortunately, the spectrum of the deflector did not allow us to do the same and therefore we used the photometric results to infer its redshift. We plot here the histogram of the spectroscopic redshifts from SDSS-DR14 \texttt{SpecPhotoAll} galaxies with similar Petrosian magnitudes (within 2$\sigma$ uncertainties).}
\label{fig:WG0214_spectra}
\end{figure}

The extracted spectra for the two position angles are shown in the upper panel of Figure~\ref{fig:WG0214_spectra}, where we label the main quasar emission lines (Ly-$\alpha$, SiIV, OIV], HeII, OIII]) that allowed a secure determination of the redshift of the source (z$_{s}=3.229\pm0.004$). This value is in good agreement (within 2--$\sigma$ error) with the value of z$_{s}=3.242\pm0.005$ reported in \cite{Lee19}.

Unfortunately the quality of the longslit spectra and the high contamination by the quasar did not allow us to securely determine the redshift of the deflector (z$_{l}$), although we identified some absorption lines (corresponding to more than one redshift, and therefore indicating the presence of galaxies along the line-of-sight).  
Therefore we infer z$_{l}$ from photometry.  
To estimate the photometric redshift of the deflector we compared its PSF colors as derived in Section~\ref{sec:photometry} with galaxy spectra from the Sloan Digital Sky Survey (SDSS, DR14, \citealt{Abolfathi18}). In particular, we select all galaxies in \texttt{SpecPhotoAll} with magnitudes within 2$\sigma$ of those obtained from our photometry (we use the \texttt{petroMag} and its \texttt{petroMagErr} for each band) of the lens, and retain their spectroscopic redshifts. The resulting histogram, plotted in the low panel of  Fig.~\ref{fig:WG0214_spectra}, is double peaked. 
We then calculate the mean and standard deviation of the both peaks, fitting two normal distributions: $z_{l_1,WG0214}=0.22\pm0.09$, $z_{l_2,WG0214}=0.53\pm0.08$. 
We note that the second peak (higher) of the histogram is consistent with the tentative redshift reported in \citet{Lee19}.

\subsection{WG2100-4452}
For WG2100-4452 we obtained only one slit position with PA = 40 degrees and width=$3.0\arcsec$, observed under thin cloud and very high high lunar brightness  (94\%, corresponding to a moon distance of 30 degrees). Unfortunately, with such a configuration, all four images of the QSO are blended in the slit; it is therefore impossible to properly separate the components, especially for the two brightest ones. For the dim C and D components, unfortunately the short integration time that we obtained (1200 sec) did not allow us to identify their signal. 
However, having clear evidence for emission lines from A+B, given the very specific geometrical configuration and the colors, and finally, given that this system can be perfectly fit with a simple lens model (as we will show in Sec~\ref{sec:models}), we are fully confident about its lensing nature. 

The accurate astrometric information provided by our analysis helped us to decompose the 2D spectrum, separating, at least partially, the lens from the source. 
To this end, we considered a two-component model, with the A+B QSO being the first and the deflector being the second. We estimated, directly from the DES images, that the separation between them is $\sim1.3\arcsec$, which corresponds to roughly $2.5$ binned pixels for our observing strategy. 

Using the Penalized Pixel Fitting (pPXF) code of \citet{Cappellari17}, we extracted the stellar velocity dispersion of the deflector from its absorption-line spectrum, using a maximum penalized likelihood approach.  
The spectrum has been extracted along an aperture of 3 pixels (corresponding to an aperture of roughly $1.5\arcsec$), after subtracting the emission from the source (the broad MgII line at $\lambda\sim5190$ \AA\ observed wavelength). 
As stellar templates for the fit, we select 100 stars (F, G, K, M) from the MILES stellar template library of \citet{Sanchez06}, which cover the wavelength range 3525–7500 \AA\ at 2.5 \AA\ FWHM spectral resolution. 
The resulting fit, yielding a $\sigma_{\star} = 287\pm40$\kms, is shown in Figure~\ref{fig:ppxf}.
The statistical uncertainty ($\delta\sigma_{stat}=38$\kms) and the systematic uncertainty ($\delta\sigma_{stat}=14$\kms due to the templates used, the region of the spectrum that we fitted, and the eighth order of multiplicative polynomial that we used to take care of the continuum) are added in quadrature. 

\begin{figure}
\includegraphics[scale=0.32,angle=0]{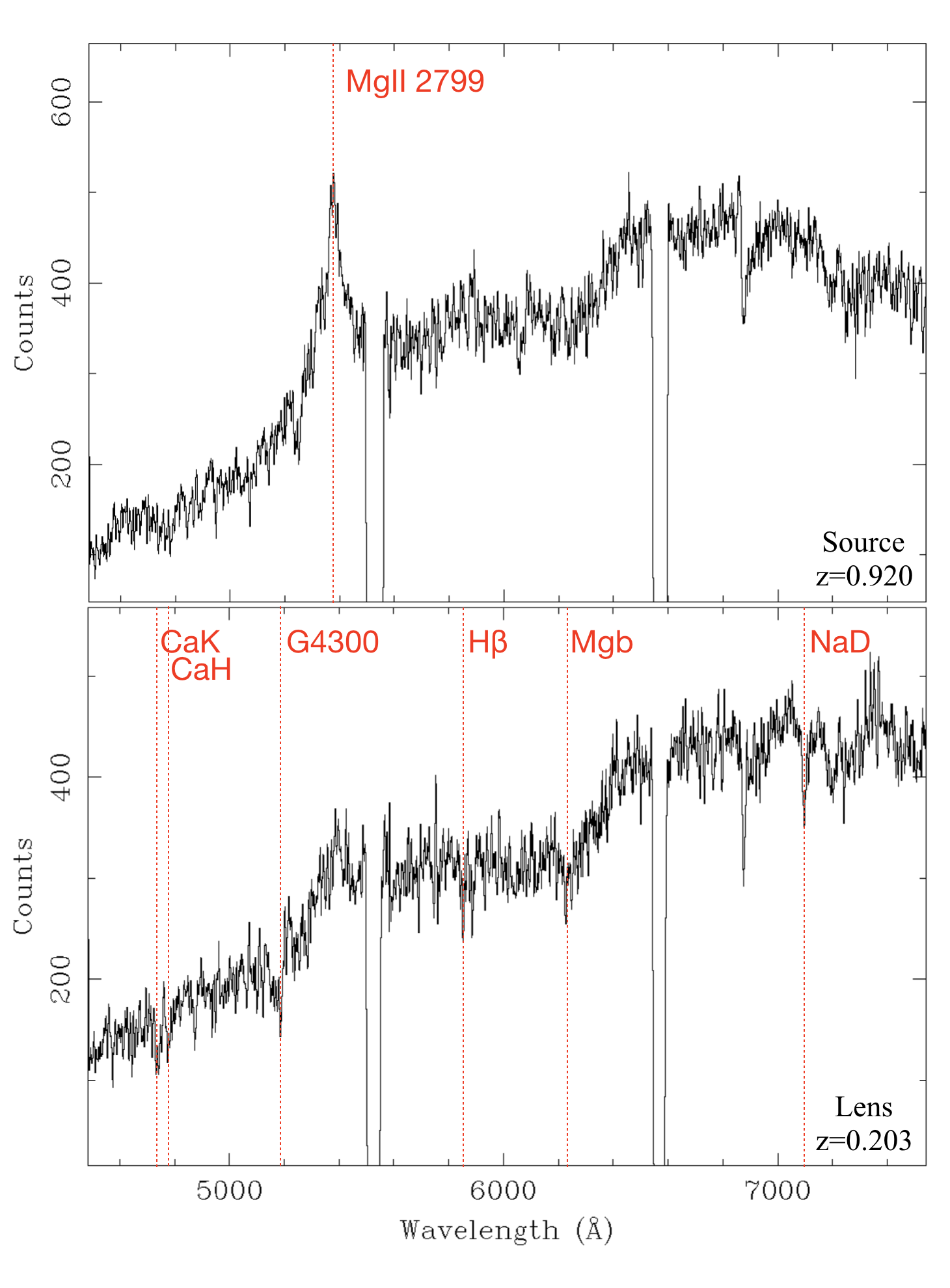}
\caption{SALT spectra of the lensed quasar (upper panel) and the deflector (lower panel) of WG2100-4452 smoothed with a boxcar of 5 pixels for better visualization. We securely infer the redshift for both the source and the deflector, thanks to the identification of emission and absorption lines, marked in the figure with vertical lines. De-blending the multiple images of the QSO is unfortunately not possible with the current slit configuration.}
\label{fig:WG2100_spectra}
\end{figure}

\begin{figure}
\begin{center}
\includegraphics[scale=0.26,angle=0]{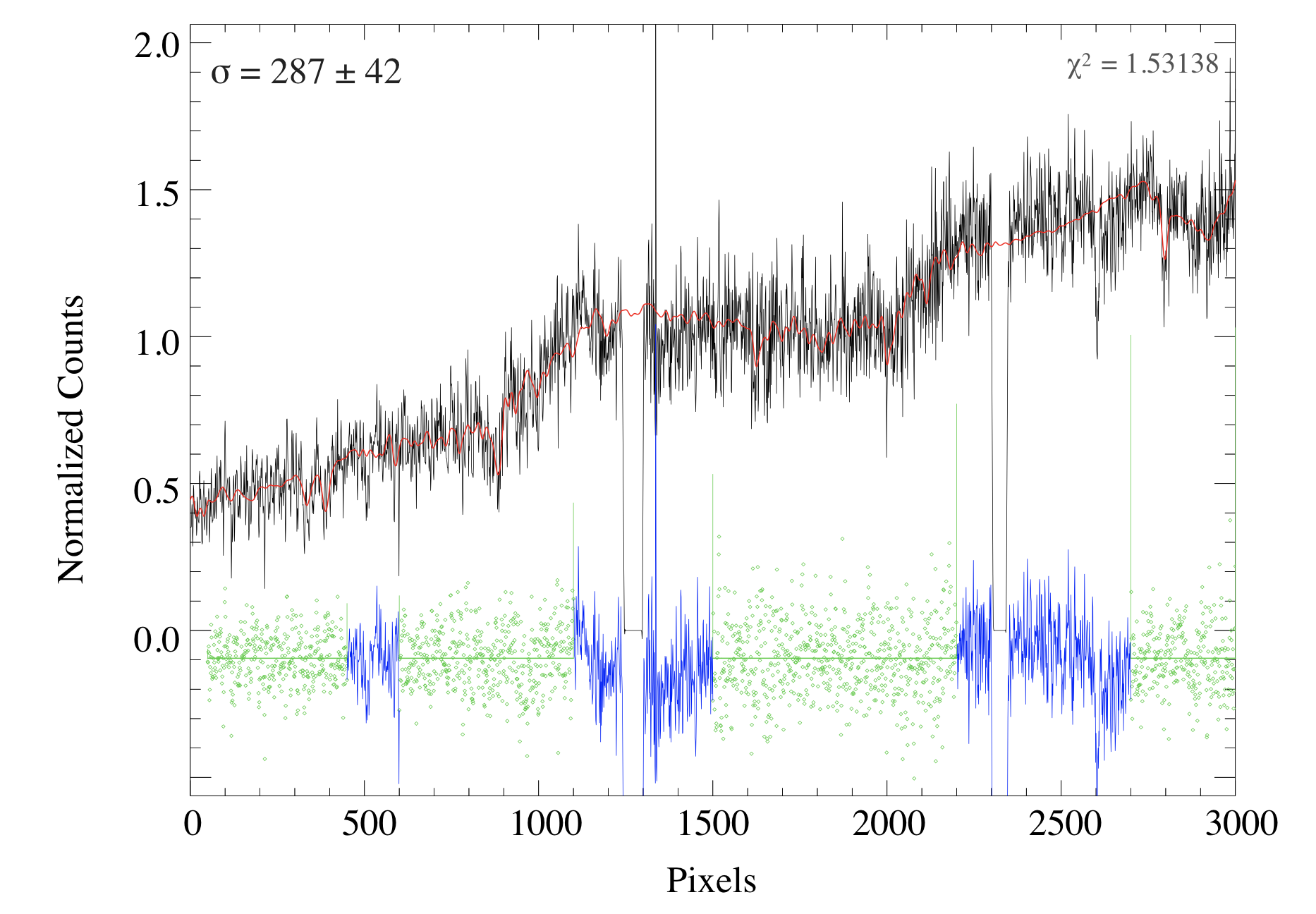}
\caption{pPXF fit of the lens galaxy in WGA2100-4452. The galaxy spectrum is plotted in black and the best-fit stellar template is overplotted in red. Residuals of the fit are shown in light green; blue regions are those excluded from the fit. }
\label{fig:ppxf}
\end{center}
\end{figure}

\section{Lens Models}
\label{sec:models}
To model the two quadruplets, we make use of the publicly available software \texttt{glafic}, presented in \citet{OguriPAS10}. 
The code allows one to efficiently fit lensed images for both point-like and extended sources, also handling multiple sources, and considering a wide range of lens potentials. 
We tested two different models to estimate the parameters: first, a Singular Isothermal Ellipsoid (SIE) mass model, and second a Power Density Ellipsoid (PDE) model where the mass density slope is left as free parameter. 

The SIE is one of the most frequently used lens model for strong gravitational lenses since it describes a mass distribution with a flat rotation curve. 
The convergence of the model is:
\begin{equation}
\kappa=\frac{\theta_{SIE}}{2\sqrt{\tilde{x}^2+\tilde{y}^2/q^2}}
\end{equation}
where $q=1-e$ is the axis ratio (with $e$ being the ellipticity), and the coordinates  $\tilde{x},\tilde{y}$ are rotated by the position angle $PA_{SIE}$, defined east of north and assuming that the positive direction is north. Finally,  $\theta_{SIE}$ is the Einstein radius, which in the spherical case is defined as
\begin{equation}
\theta_{SIE}=\frac{4\pi}{\sqrt{q}}\left(\frac{\sigma_{SIE}}{c}\right)^2\frac{D_{ls}}{D_s}
\end{equation}
where $D_{s}$ is the angular diameter distance of the source from the observer, $D_{ls}$ is the relative distance between the lens and the source and $ \sigma_{SIE}$ is the velocity dispersion of the deflector. 

The PDE model is more useful than the SIE model for studying the effect of the radial density slope on lensing. The convergence, defined in \citet{Keeton01}, is:
\begin{equation}
\kappa=\frac{3-\beta}{2} (\theta_{PDE})^{1-\beta}\left(\tilde{x}^2+\tilde{y}^2/q^2\right)^{(1-\beta)/2}
\end{equation}
where $\theta_{PDE}$ is the Einstein Radius. 
The PDE model corresponds to an isothermal model for $\beta = 2$. 

The SIE model has seven free parameters: Einstein radius, x and y coordinates of the galaxy and of the source (true position), ellipticity and position angle. The PDE has eight free parameters: Einstein radius, x and y coordinates of the galaxy and of the source, ellipticity, position angle and mass density slope.  
Since from the DES images and photometric analysis we have eight constraints for each system (x,y, apparent positions of the 4 multiple images), both models are fully constrained. 

We use the inferred spectroscopic redshifts for the sources (z$_{s,WG0214} = 3.229$, z$_{s,WG2100} = 0.92$). 
For the deflector redshifts, we used the spectroscopic value obtained for WG2100-4452 (z$_{l,WG2100} = 0.203$) and the photometric ones for WG0214-2105 (we run the the model twice, one time assuming z$_{l_1, WG0214_1} = 0.22\pm0.09$, and the second time assuming z$_{l_2, WG0214} = 0.53\pm0.08$). 
The model results are reported in Table~3, where we also report reduced chi-square ($ \chi^{2}_{\nu}$) values to provide a quantitative measure of the goodness of the fit. Overall a good agreement between the two model is found. 
We also obtained estimates of the deflectors' masses inside the Einstein radii ($M_{WG0214} = 8.52\times 10^{10}M_{\odot}$, $M_{WG2100} = 1.42\times10^{11}M_{\odot}$) \footnote{$M_{Ein} = \theta^{2}\dfrac{c^2D_{l}D_{s}}{4GD_{ls}}$, where $D_{l}$ is the distance of the lens from the observer, $D_{s}$ is the distance of the source from the observer, $D_{ls}$ is the relative distance between the lens and the source, $ \theta$ is the Einstein Radius and finally, c is the speed of light}.  
For the PDE model, the best fit mass density slopes we recover are $\beta_{WG0214} = 2.1\pm0.2$ and $\beta_{WG2100} = 1.9\pm0.2$, both consistent with an isothermal profile. 

Uncertainties in model parameters were calculated using a Markov-Chain Monte-Carlo (MCMC) approach. We generated 100000 models and obtained a normal distribution for each of the estimated parameters, then computing the mean and standard deviation values.  

\begin{table*}
\center
\label{tab:lensing_model}
\caption{Model parameters obtained for the two quadruplets: ellipticity, position angle and Einstein radius from both models; mass density slope from PDE and galaxy velocity dispersion from SIE. We also report, for each model and each system, the modest value of the $\chi^2$ divided by the degrees of freedom to judge the quality of the fit. }
\begin{tabular}{|l|c|c|c|c|c|c|c|c|c|c|}
\hline 
\hline
ID &  $e_{SIE}$ & $e_{PDE}$ &  $PA_{SIE}$ & $PA_{PDE}$ & $\theta_{SIE}$ & $\theta_{PDE}$ & $\beta_{PDE}$ & $\sigma_{SIE}$ & $\chi^2_{\nu}$ & $\chi^2_{\nu}$ \\ 
   & & &  (deg) &  (deg) & (") & (") &  & (kms$^{-1}$)   & SIE &  PDE \\ 
\hline 
WG0214-2105 &  $0.3\pm0.1$ & $0.34\pm0.16$ & $62.5\pm1.1$ & $62.5\pm1.1$ & $0.89\pm0.02$ & $0.91\pm0.02$ & $2.1\pm0.2$ & $211\pm2$  &  1.16 & 1.07 \\
\hline
WG2100-4452 & $0.39\pm0.05$ & $0.34\pm0.15$ & $0.4\pm0.7$ & $0.6\pm1.0$ & $1.30\pm0.02$ & $1.32\pm0.03$ & $1.9\pm0.2 $  & $251\pm2$ & 1.08 & 0.96 \\
\hline
\hline

\end{tabular}
\end{table*} 



\begin{figure*}
\includegraphics[scale=0.43,angle=0]{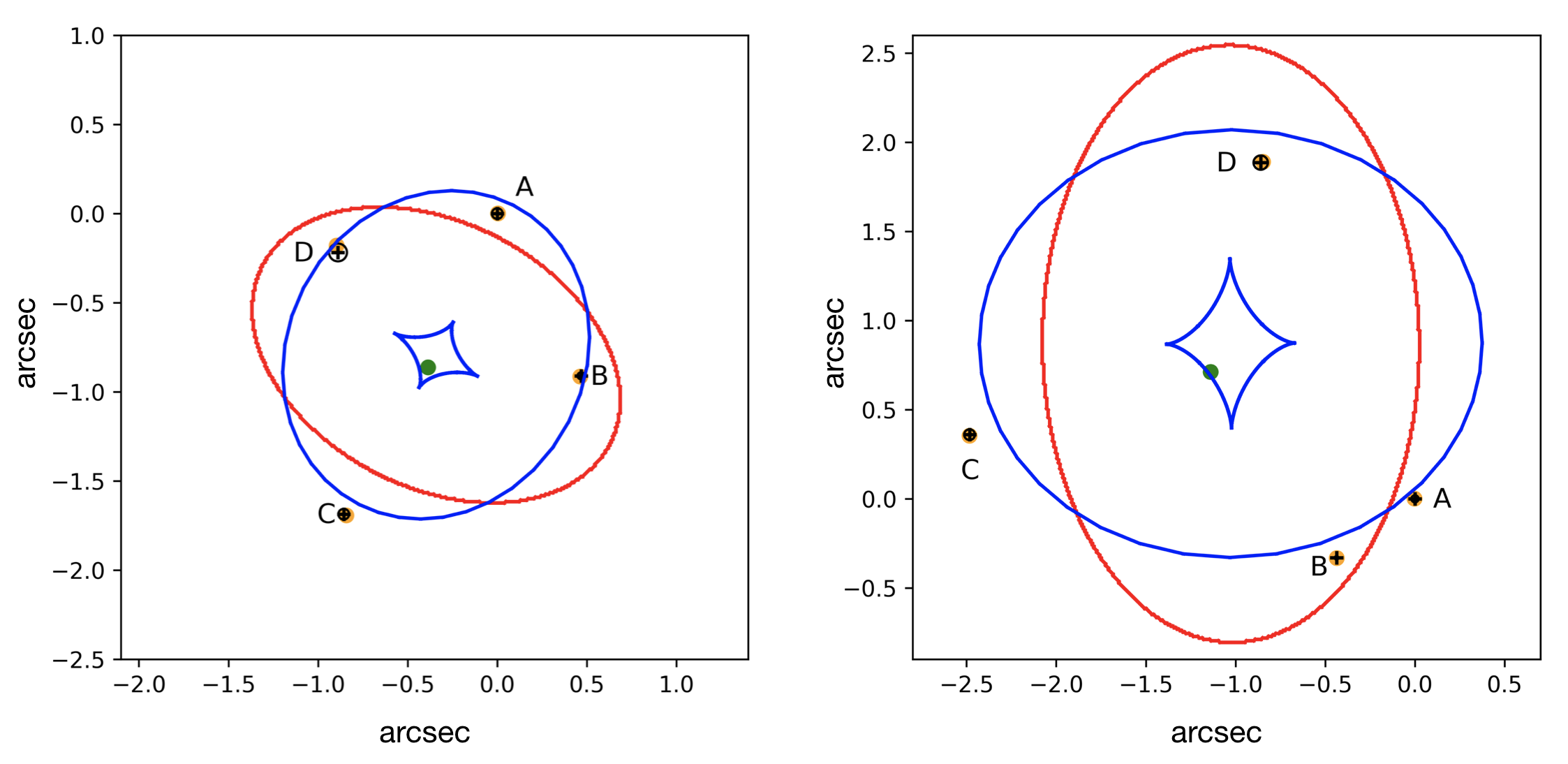}
\caption{Mass model fit obtained assuming a SIE profile matter distribution for WG0214-2105 (left panel) and WG2100-4452 (right panel). The red and blue curves corresponding to the critical curves and caustics of the model, respectively. The black circles with internal crosses show the observed positions inferred from the DES images, the orange filled circles show the predicted image positions and finally the green filled circle indicates the calculated true source position for each system. The QSO components are labeled following the same convention adopted in Figure~\ref{fig:psf}. }
\label{fig:Mass_mod}
\end{figure*}

Figure~\ref{fig:Mass_mod} shows the mass models for WG0214-2105 (left panel) and WG2100-4452 (right panel) in the SIE case. We plot caustics and critical curves, as well as the apparent and true positions of the sources.  
We note that the flux ratios could be strongly perturbed by microlensing effects.  
Therefore, the magnitude difference of the quasar images has been excluded from the model.

Finally, we calculated the magnification and time delay values for all components, relative to the brightest one (A), and report the results in Table~4 for both models. 
We caution the reader that microlensing can influence time-delay measurements, making them time variable.  This is due, as described in \citet{Tie18}, to a combination of two effects. First, the quasar disk has an inclination with respect to the line-of-sight, and thus different parts of the disk lie at different projected distances to the source. This configuration can change time delays on the scale of the light crossing time of the accretion disk (typically of the order of light days). Second, microlensing can cause a differential magnification of the temperature of the emission from the disk, causing a differential time delay on the scale of the light crossing time. 
Moreover, we note that the time-delays highly depend on the deflector redshift: increasing the galaxy redshift, the time-delay values increase proportionally.

The relative positions inferred from the lensing model and directly calculated from the images (in $r$-band), always with respect to the brightest QSO component (A), agree perfectly with each other. This is also clearly visible from Figure~\ref{fig:Mass_mod}, where we report with orange filled circles the model predicted positions and with empty black circled crosses the positions directly obtained from the DES images.

\begin{table*}
\center
\label{tab:timedel}
\caption{Quasar image parameters for the best fits obtained for the two models we tested: a Singular Isothermal Ellipsoid (SIE) mass model and a Power Density Ellipsoid (PDE) model. Component A is taken as reference. Since the time-delays depend on the deflector redshift, for WG0214-2105, we provide time-delay values for both the fiducial photometric redshifts ($z_{l_1,WG0214}=0.22\pm0.09$, $z_{l_2,WG0214}=0.53\pm0.08$) separated by commas. }
\begin{tabular}{|c|c|c|c|c|c|c|c|}
\hline 
\hline
Name & Comp. & $\delta$x & $\delta$y & $\Delta{mag}_{SIE}$ & $\Delta{t}_{SIE}$ & $\Delta{mag}_{PDE}$ & $\Delta{t}_{PDE}$ \\
 &  & (") & (") & (mag) & (days) & (mag) & (days) \\
\hline 
WG0214-2105 & A & 0.003 & 0.001 & $\equiv 0$ & $\equiv 0$ & $\equiv 0$ & $\equiv 0$  \\
WG0214-2105 & B & 0.462 & -0.914 & -0.72 & 0.9, 3.5& -0.7 & 1.2, 3.2 \\
WG0214-2105 & C & -0.842 & -1.693 & -0.75 & -4.5, -12.5 & -0.6 & -4.1, -11.3  \\
WG0214-2105 & D & -0.898 & -0.179 & -0.56 & 1.3, 2.5 & -0.53 & 0.8, 2.5 \\
\hline 
WG2100-4452 & A & -0.003 &0.002 & $\equiv 0$ & $\equiv 0$ & $\equiv 0$ & $\equiv 0$ \\
WG2100-4452 & B & -0.437 & -0.331 & -0.12 & 0.2  & 0.00 & 0.1 \\
WG2100-4452 & C & -2.484 & 0.354 & -1.29 & -9.3  & -1.37 & -4.0 \\
WG2100-4452 & D & -0.847 & 1.891 & -2.0 & 11.6  & -1.79 & 6.1 \\
\hline 
\hline
\end{tabular}
\end{table*} 

\section{Conclusions}
\label{sec:conclusions}
In this paper, we have presented the spectroscopic confirmation and lens modelling of two new quadruply lensed quasars recently found in the DES public footprints, namely WG0214-2105 and WG2100-4452. 
The quadruplets were both found with a method based on three main steps: an infrared (WISE) color-preselection of QSO-like objects, a morphological criterion based on multiple matches in the Gaia DR2 catalog and finally, visual inspection of the outcomes. 
We refer the reader to A18 and \citet{Spiniello18} for a detailed description of the search methodology and to A18 and ARN18 for the discovery report. 

Since the only missing ingredient to unambiguously confirm the lensing nature of these objects, whose geometrical configuration and chromaticity are typical of lensed QSOs, was spectroscopy, we targeted them with the Southern African Large Telescope. 
For both systems, we calculated the redshifts of the deflectors and the sources, confirming their multiple lensing nature. 
We obtained $z_{\rm source} = 3.229\pm0.004$ for WG0214-2105 thanks to the identification of the strong Ly-$\alpha$ line and other weaker lines such as SiIV, OIV], HeII, OIII].  We inferred instead $z_{\rm source} = 0.920\pm0.002$
for WG2100-4452, thanks to the identification of the prominent MgII emission line. 
Only for WG2100-4452 we could estimate the deflector redshift directly from the spectral absorption lines (i.e. CaK, CaH, G4300, H$\beta$, Mgb,NaD), since the galaxy is quite bright ($r=17.68\pm0.01$ mag, see Table~2).  For the other lens, we infer the redshift using optical magnitudes that we obtained performing Direct Image Analysis (DIA) and photometry (presented in Sec.~\ref{sec:photometry}).  

Microlensing studies require multiple images of a system taken in different epocs. We presented here preliminary evidence of time variability for one of the two systems, WG0214-2105, obtained measuring the photometry of all the components from DES (taken in 2016) and Pan-STARRS (2014).  
A change of $\approx 0.4$ mag is visible for the faintest QSO component (D), and other two components show smaller changes too. Although with only two epochs  separated by two years, intrinsic chromatic variability cannot be excluded, we argue in favor of microlensing because we do not see any clear correlation with wavelength. Moreover, the lens model inferred time-delays are of the order of a few days and the probability that a quasar changes its magnitude by $\approx$ half a magnitude in such a short time is very low.
We also note that, as shown in \citet{Schechter2002}, microlensing in saddlepoints (B, D) is more likely in the presence of a smoothly distributed (dark matter) component.

Finally, we confirm the work of \citet{Wynne18}, finding that these systems are perfectly compatible with simple lens models. 
We tested two different lens models, specifically a Singular Isothermal Ellipsoid approximation where the mass density slope is fixed but the stellar velocity dispersion is allowed to vary, and a Power Density Ellipsoid model where the Einstein radius and the mass density slope are free parameters. For both models, we constrain the x and y positions of the source and the deflector, the ellipticity and the position angle. 
We finally obtained for both quadruplets changes in magnitudes and time-delay measurements for each of the quasar multiple images, finding that while the former do not depend on the chosen model, the latter heavily depend on it. 

We thus conclude that the planned integration times (set by the magnitudes and colors of the sources in the photometric survey catalogs) are long enough to obtain SALT spectra with sufficient signal-to-noise ratios to detect emission lines from the QSOs and, in most of the cases, also absorption lines from the lens. 
Our observing strategy works properly and will allow us, once the program is completed, to confirm many among the $\sim300$ lens candidates (arcs and QSOs) that we selected in the last two years from the KIDS and KABS surveys.

\section*{Acknowledgments} 
CS has received funding from the European Union's Horizon 2020 research and innovation programme under the Marie Sk\l odowska-Curie actions grant agreement No 664931. 
LM and MV acknowledge support by the Italian Ministry of Foreign Affairs and International Cooperation (MAECI Grant Number ZA18GR02) and the South African Department of Science and Technology’s National Research Foundation (DST-NRF Grant Number 113121) as part of the ISARP RADIOSKY2020 Joint Research Scheme. 
NRN acknowledges financial support from the European Union's Horizon 2020 research and innovation programme under the Marie Sk\l odowska-Curie grant agreement No 721463 to the SUNDIAL ITN network. LVEK and GV are supported through an NWO-VICI grant (project number 639.043.308). Finally SS acknowledges support from the STFC grant ST/P000584/1.

All of the observations reported in this paper were obtained with the Southern African Large Telescope (SALT). 
This publication also makes use of data products from the Wide-field Infrared Survey Explorer, which is a joint project of the University of California, Los Angeles, and the Jet Propulsion Laboratory/California Institute of Technology, funded by the National Aeronautics and Space Administration. 
The Pan-STARRS1 Surveys (PS1) and the PS1 public science archive have been made possible through contributions by the Institute for Astronomy, the University of Hawaii, the Pan-STARRS Project Office, the Max-Planck Society and its participating institutes, the Max Planck Institute for Astronomy, Heidelberg and the Max Planck Institute for Extraterrestrial Physics, Garching, The Johns Hopkins University, Durham University, the University of Edinburgh, the Queen's University Belfast, the Harvard-Smithsonian Center for Astrophysics, the Las Cumbres Observatory Global Telescope Network Incorporated, the National Central University of Taiwan, the Space Telescope Science Institute, the National Aeronautics and Space Administration under Grant No. NNX08AR22G issued through the Planetary Science Division of the NASA Science Mission Directorate, the National Science Foundation Grant No. AST-1238877, the University of Maryland, Eotvos Lorand University (ELTE), the Los Alamos National Laboratory, and the Gordon and Betty Moore Foundation.

\end{document}